\documentstyle[epsfig,12pt]{article}
\begin{document}
%
%
\newcommand{\x}{\cdot}
\newcommand{\ra}{\rightarrow}
\newcommand{\pom}{\mbox{${\rm \cal P}$omeron}}
\newcommand{\reg}{\mbox{${\rm \cal R}$eggeon}}
\newcommand{\flux}{\mbox{$F_{{\cal P}/p}(t, \xi)$}}
\newcommand{\ap}{\mbox{$\bar{p}$}}
\newcommand{\pap}{\mbox{$\bar{p} p$}}
\newcommand{\SPS}{\mbox{S\pap S}}
\newcommand{\xp}{\mbox{$x_{p}$}}
\newcommand{\sumet}{\mbox{$\Sigma E_t$}}
\newcommand{\mpr}{\mbox{${m_p}$}}
\newcommand{\mpi}{\mbox{${m_\pi}$}}
\newcommand{\rs} {\mbox{${\sqrt {s}}$}}
\newcommand{\rsp}{\mbox{$\sqrt{s'}$}}
\newcommand{\rsps}{\mbox{$\sqrt{s} = 630 $ GeV}}
\newcommand{\lum}{\mbox{$\int {\cal L} {dt}$}}
\newcommand{\T}{\mbox{$t$}}
\newcommand{\abt}{\mbox{${|t|}$}}
\newcommand{\di}{\mbox{d}}
\newcommand{\HS}{\mbox{$xG(x)=6x(1-x)^1$}}
\newcommand{\sigdifjets}{\mbox{$\sigma_{sd}^{jets}$}}
\newcommand{\sigpomjets}{\mbox{$\sigma_{{\cal P}p}^{jets}$}}
\newcommand{\sigdiftot}{\mbox{$\sigma_{sd}^{total}$}}
\newcommand{\sigpomtot}{\mbox{$\sigma_{p {\cal P}}^{total}$}}
\newcommand{\sigpptot}{\mbox{$\sigma_{pp}^{total}$}}
\newcommand{\sigpomzero}{\mbox{$\sigma_{{\cal P}p}^o$}}
\newcommand{\dsig}{\mbox{${d^2 \sigma }\over{d \xi dt}$}}
\newcommand{\alamb}{\mbox{$\overline{\Lambda^{\circ}}$}}
\newcommand{\lamb}{\mbox{$\Lambda^{\circ}$}} 
\newcommand{\peetee}{\mbox{${ p_t}$}}
\newcommand{\PRET}{\mbox{\Proton-\sumet}}
%
\newcommand{\xpom}  {\mbox{$x_{I \! \! P}$}} 
\newcommand{\pmm}         {\mbox{${\cal P}$}}
\newcommand{\gm}         {\mbox{$\gamma^{*}$}}
\newcommand{\gmp}         {\mbox{$\gamma^{*} p$}}
\newcommand{\siggp}      {\mbox{$\sigma_{\gamma^{*} p}^{total}$}}
\newcommand{\siggpm}     {\mbox{$\sigma_{\gamma^{*} {\cal P}}^{total}$}}
\newcommand{\FtwoDtwo}       {\mbox{$F_2^{D(2)}$}}
\newcommand{\FtwoDthree}       {\mbox{$F_2^{D(3)}$}}
\newcommand{\xpomFtwo}     {\mbox{$\xi \FtwoDthree$}}
\newcommand{\qsq}        {\mbox{$Q^2$}}
\newcommand{\mx}         {\mbox{$M_X$}}
\newcommand{\mxsq}       {\mbox{$M_X^2$}}
\newcommand{\w}          {\mbox{$W$}}
\newcommand{\wsq}        {\mbox{$W^2$}}
\newcommand{\fluxint}    {\mbox{$f_{{\cal P}/p}(\xi)$}}
\newcommand{\eps}        {\mbox{$\epsilon$}}
\newcommand{\alfp}        {\mbox{$\alpha '$}}
\newcommand{\alf}        {\mbox{$\alpha $}}
\begin{titlepage}
\vspace{4cm}
\begin{flushright}  
{12 March, 2000}
\end{flushright}

\vspace{4ex}
\begin{center}
{
\LARGE
\centerline{\bf \boldmath Inelastic diffraction data}
\centerline{\bf \boldmath and the effective \pom\ trajectory}
\normalsize
}
\vspace{11 ex}
Samim Erhan\footnote{samim.erhan@cern.ch} and 
Peter Schlein\footnote{peter.schlein@cern.ch} \\
\vspace{3.0mm}
University of California\footnote{Supported by U.S. National Science Foundation
Grant PHY94--23142}, Los Angeles, California 90095, USA. \\
\end{center}
\vspace{13 ex}
\begin{abstract}
A further analysis of inelastic diffraction data at the ISR and SPS-Collider 
confirms the relatively flat $s$--independent \pom\ trajectory in the 
high--$|t|$ domain, $1 < |t| < 2$~GeV$^2$, reported earlier by the UA8 
Collaboration. 
At $|t| = 1.5$~GeV$^2$, $\alpha = 0.92 \pm 0.03$ is in agreement with the 
trajectories found in diffractive photoproduction of vector mesons at HERA.
This suggests a universal fixed \pom\ trajectory at high--$|t|$.
We also show that a triple--Regge \pom --exchange parametrization fit to the 
data requires an $s$--dependent (effective) \pom\ trajectory intercept, 
$\alpha (0)$, which decreases with increasing $s$,
as expected from unitarization (multi--\pom --exchange) calculations.
$\alpha (0)$ = 1.10 at the lowest ISR energy, 1.03 at the SPS--Collider 
and perhaps smaller at the Tevatron. 

\end{abstract} 
\vspace{6 ex}
\begin{center}
in press: Physics Letters B \\
\end{center}
\vspace{6 ex}
\end{titlepage}
\setlength{\oddsidemargin}{0 cm}
\setlength{\evensidemargin}{0 cm}
\setlength{\topmargin}{0.5 cm}
\setlength{\textheight}{22 cm}
\setlength{\textwidth}{16 cm}
\setcounter{totalnumber}{20}
\clearpage\mbox{}\clearpage
\pagestyle{plain}
\setcounter{page}{1}
\section{Introduction}
\label{sect:intro}
\indent

\begin{figure}
\begin{center}
\mbox{\epsfig{file= 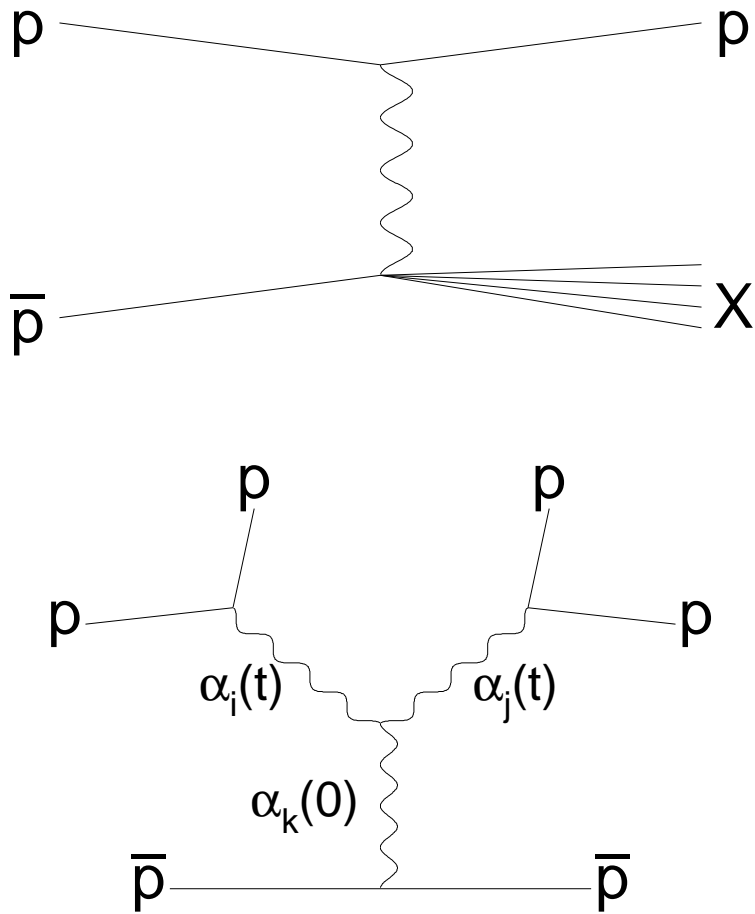,width=16cm}}
\end{center}
\caption[]{
Upper: The diffractive \pap\ process. 
The exchanged \pom\ has a momentum transfer, \T , and 
momentum fraction, $\xi \equiv \xpom = 1 - \xp$, of the
incident proton. The squared invariant mass of the $X$ system is 
$M_X^2 = s' = \xi s$. Lower: The triple--Regge version of the
upper process.
}
\label{fig:diag}
\end{figure}

Single diffraction, or
the inclusive inelastic production of beam-like particles
with momenta within a few percent of the associated incident beam 
momentum, as in:
\begin{equation}
p (\ap) \, \,  \, \, + \, \, p_i \, \,  \, \, \ra \, \, \, \, X \, \, 
+ \, \, p_f 
\label{eq:dif}
\end{equation}
has been studied for more than 30 years.
The chief characteristic of data from these processes is the existence of
a pronounced enhancement at Feynman-$x_p$ of $p_f$ near unity,
with the absence of other particles nearby in rapidity (``rapidity gap").
This is interpreted using Regge 
phenomenology[1--6] as evidence for the dominance of color-singlet
\pom --exchange (see Fig.~\ref{fig:diag}).
The observed $x_p$ spectrum 
reflects the distribution of the exchanged \pom 's momentum fraction
in the proton\footnote{We use the symbol $\xi$ for this variable in view
of its simplicity and its increasing use in the literature.}, 
$\xi \equiv x_{I \! \! P} = 1 - \xp$. 

A relatively recent idea\cite{is,berger,dl_hard} underlying the 
phenomenology is 
that, although the \pom 's existence in the proton is due to non-perturbative
QCD, once the \pom\ exists, perturbative QCD processes can occur 
in proton-\pom\ and $\gamma^{*}$-\pom\ interactions.
Ref.~\cite{is} proposed the study of 
such hard processes in order to determine the \pom\ structure.
First ``hard diffraction" 
results were obtained by the UA8 collaboration\cite{ua8}
using React.~\ref{eq:dif}, and by the H1\cite{h1} and ZEUS\cite{zeus} 
collaborations using $e p$ interactions. 
Hard diffraction results on React.~\ref{eq:dif} also exist from
the CDF\cite{hardcdf} and D0\cite{hardd0} collaborations at the Tevatron.

Factorization of \pom\ emission and interaction in the 
inclusive React.~\ref{eq:dif} is expressed by writing the 
single-diffractive differential cross section as a 
product of a ``Flux Factor'' of the \pom\ in the proton, \flux , 
and a proton-\pom\ total cross section (see Sect.~\ref{sect:fits}):
\begin{equation}
{{d^2 \sigma_{sd}}\over{d \xi dt}} \, \, 
= \, \,  \flux \, \, \x \, \, \sigpomtot(s')
\label{eq:factorhad}
\end{equation}
$s'$ is the squared invariant mass of the $X$ system and,
to good approximation, is given by: $s' = \xi s$.
\T\ is the \pom 's four--momentum transfer.
There are many examples in the literature of the validity of factorization
(see for example Refs.~\cite{ua8,h1,zeus,ua8dif,cool}), 
and our working assumption in the present
paper is that Eq.~\ref{eq:factorhad} is a good approximation. 
There is, however, a long--standing unitarity problem with the
\pom --exchange prediction for React.~\ref{eq:dif} which deserves
re--examination.

\begin{figure}
\begin{center}
\mbox{\epsfig{file=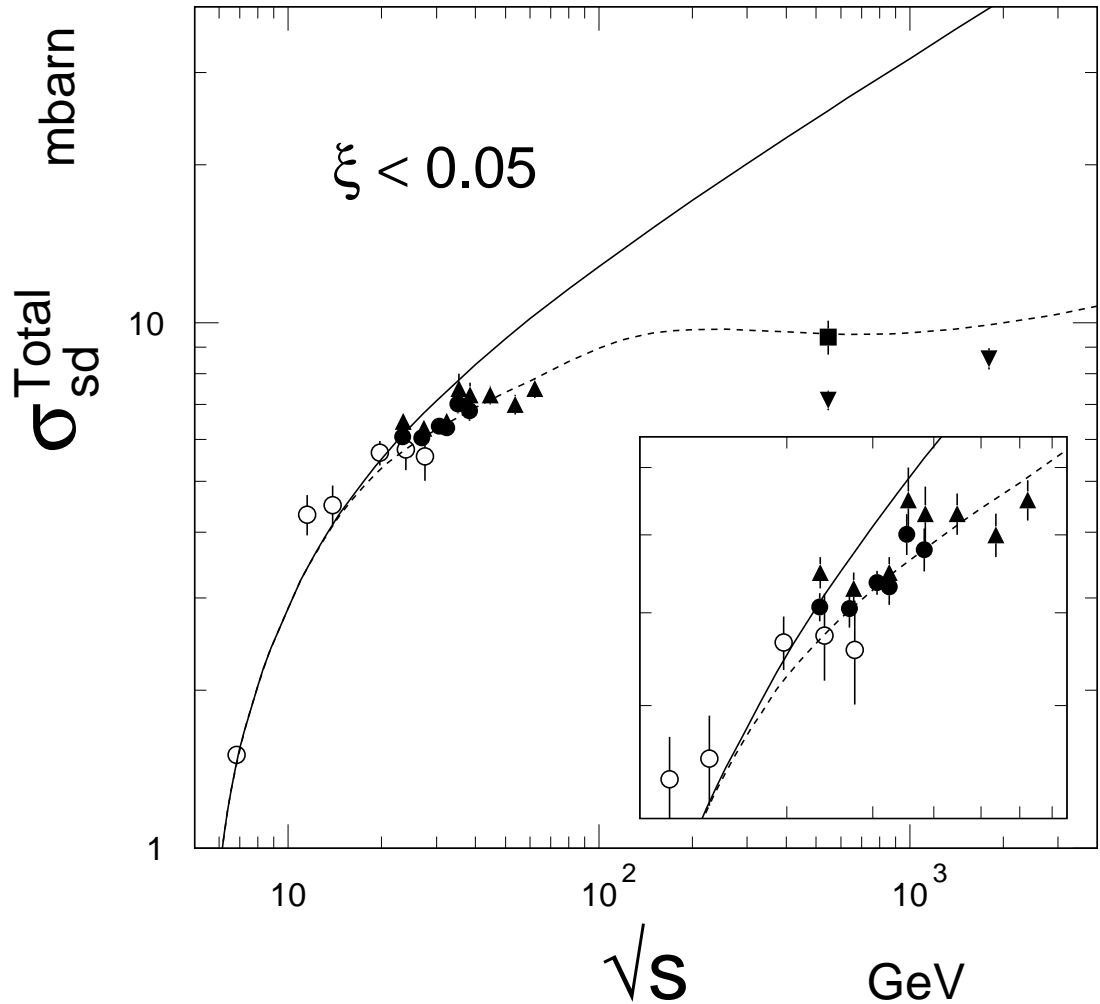,width=16cm}}
\end{center}
\caption[]{
Measured total single diffractive cross section for $\xi < 0.05$ in $pp$ or 
$p\ap$ interactions vs. $\sqrt{s}$.
A factor of two is included to account for
both hemispheres (see the experiment references in Ref.~\cite{es1}).
The insert is a blow--up of the ISR energy range.
The solid curve is the UA8 Triple--Regge prediction;
the dashed curve shows the consequence of 
multiplying it by a ``toy" damping factor\cite{es1}.
}
\label{fig:sigtot}
\end{figure}

The rising total cross sections observed at Serpukhov ($K^+ p$) and
the ISR ($pp$) in the early 1970s
led to the conclusion\cite{cgm,ck,ttwu} that the effective
\pom\ Regge trajectory intercept at \T\ = 0, 
$\alpha (0) = 1 + \eps $, was larger than 
unity\footnote{The fit result for the trajectory in Ref.~\cite{cgm} was 
1.06 + 0.25\T ; the latest refined fits\cite{cudell,dino2,dl_tot} to the
$s$--dependences of all total cross sections yield $\epsilon \sim 0.10$. }.
Although this violates the Froissart-Martin unitarity 
bound\cite{froissart,martin} for total cross sections, 
it presents no difficulty at present and forseeable collider energies.
However, this is not the case for partial cross sections such as diffraction.
This is easily seen by examining the dominant 
$\xi$--dependent Regge factor in Eq.~\ref{eq:factorhad} at small--$\xi$ and 
small--$t$:
\begin{equation} 
\flux \, \, \propto \, \, \xi^{1-2 \alpha(0)} \, \, = \, \, 
{{1}\over{\xi^{1 + 2 \epsilon}}}, 
\label{eq:xiinF}
\end{equation}
Kinematically, $\xi$ has a minimum value in React.~\ref{eq:dif},
$\xi_{min} = s'_{min}/s$, which decreases with increasing energy, such that
the rise in \flux\ at small $\xi$ becomes more and more pronounced.
With \eps\ = 0.10,
this leads to a rapidly increasing predicted total single diffractive cross 
section, \sigdiftot , with $s$, shown as the solid curve in 
Fig.~\ref{fig:sigtot} \cite{es1}.
Of course, the observed \sigdiftot\ does not display this behavior, but 
rises much more 
modestly\footnote{For reasons explained below in Sect.~\ref{sect:cdf},
we tend to discount the smaller of the two \sigdiftot\ values
at \rs\ = 546 GeV\cite{cdf}.} with $s$.

This discrepency between the predictions and the observed 
\sigdiftot\ should be understandable in the framework of Gribov's Reggeon 
calculus\cite{gribov} through multi--\pom --exchange effects (Regge cuts), 
described variously in the literature as screening, shadowing, absorption or 
damping \cite{ckt,kpt,abramovsky,glm3}.
Eq.~\ref{eq:factorhad}, traditionally used with the \eps\ obtained from 
fitting to the $s$--dependence of
total cross section data, does not take these effects into account.

It is expected that multi--\pom --exchange effects increase with $s$. This 
corresponds to a decreasing effective \eps\ \cite{kpt}
which will suppress \sigdiftot\ corresponding to the observed behavior.
To the best of our knowledge, effective \eps\ values have 
never been directly extracted from \sigdiftot\ data on React.~\ref{eq:dif}, 
although Schuler and 
Sj\"{o}strand\cite{ss} have developed a model of hadronic diffractive
cross sections in which they use $\eps = 0$ as a reasonable approximation
at the highest energies.

It was also suggested~\cite{ckmt,predazzi,glm2} that the observation 
in \gmp\ interactions at HERA, of an $Q^2$-dependent effective \pom\ 
intercept~\cite{h1varQsq,zeusvarQsq,allm}
could be the result of the decrease of screening with increasing $Q^2$.

In the present paper we use the measured \sigdiftot\ values to determine the 
effective \eps\ values as a function of energy.
We then fit to the \T --dependence of $d\sigdiftot / dt$ ($\xi < 0.05$) at the 
ISR and SPS--Collider to obtain more reliable values of \eps\ as well as the 
slope, $\alpha '$ at \T\ = 0. These latter fits also provide confirming
evidence of the relatively flat $s$--independent trajectory 
in the higher--$|t|$ region, 1.0--2.0 GeV$^2$, which was previously
reported by the UA8 Collaboration\cite{ua8dif}. 
The value of this trajectory at $|t| = 1.5$~GeV$^2$ is consistent with the 
trajectory obtained from photoproduction of vector mesons at HERA.

We note that a decreasing effective intercept has the effect of
suppressing the event yield at small--$\xi$ {\it and} small--$|t|$, 
as suggested by the data\cite{es1}.
In Ref.~\cite{es1}, we introduced an ad hoc Damping Factor to 
account for the observed suppression of the total diffractive cross section.
However, since it was observed that the damping effects extend to 
larger $\xi$ at the largest Tevatron energy,
our fixed Damping Factor had limited applicability.
In the present paper, we show that the $s$--dependent effective intercept,
as a manifestation of multi--\pom --exchange, offers a physics 
explanation for the effect and seems to be applicable up the the highest
available energies. 

Sect.~\ref{sect:fits} summarizes the analysis by the 
UA8 collaboration\cite{ua8dif}, in which they
fit Eq.~\ref{eq:factorhad} to ISR and SPS data; 
they obtain parametrizations of \flux\ and \sigpomtot\ which embody
features not previously known and specify the \pom\ trajectory at high--$|t|$. 
Sect.~\ref{sect:sdep} shows how the effective \pom\ intercept depends on 
interaction energy in React.~\ref{eq:dif} and how predictions at
\rs\ = 1800 GeV agree with the CDF collaboration's results\cite{cdf}.
The analysis in Sect.~\ref{sect:traj} yields a new \pom\ trajectory which 
depends on $s$ only at low--$|t|$. 
Finally, Sect.~\ref{sect:conclude} contains our conclusions and a discussion 
of some consequences.

\section{UA8 Triple--Regge fits and the {\boldmath \pom } trajectory at 
high--{\boldmath $|t|$ }}
\label{sect:fits}
\indent

The UA8\cite{ua8dif} collaboration analyzed data from their experiment at the 
CERN SPS--Collider (\rs\ = 630 GeV) in the $|t|$--range, 0.90--2.00~GeV$^2$,
and from the CHLM experiment at the CERN ISR\cite{albrow} (\rs\ = 23--62~GeV)
in the $|t|$--range, 0.15--2.35~GeV$^2$. 

Eq.~\ref{eq:factorhad} was fit to the data, using the dominant two terms in 
the Mueller--Regge expansion \cite{mueller,ellissanda,ffrr},  
${\cal P}{\cal P}{\cal P}$ and ${\cal P}{\cal P}{\cal R}$ 
(see Fig.~\ref{fig:diag}), 
for the differential cross section of React.~\ref{eq:dif}.
These correspond, respectively, to \pom\ exchange and the exchange of other 
non--leading, 
C=+ \reg\ trajectories (e.g., $f_2$) in the proton--\pom\ 
interaction\footnote{In reality, the $(s')^{\epsilon}$ terms have the
form $(s'/s_0)^{\epsilon}$ with $s_0 = 1$~GeV$^2$.)}, 
\begin{equation}
{{d^2 \sigma_{sd}}\over{d \xi dt}} 
\, = \, [K \, F_1(t)^2  \, e^{bt} \, \xi^{1-2\alpha (t)}] 
\, \x \, \sigma_0 [(s')^{\epsilon_1} \, + \, R \, (s')^{\epsilon_2}].
\label{eq:tripleP2}
\end{equation}
Comparing with Eq.~\ref{eq:factorhad},
the left--hand bracket is the \pom\ flux factor, \flux ,
and the right--hand bracket (together with $\sigma_0$) is 
the proton--\pom\ total cross section, \sigpomtot . 
Because this expression for \sigpomtot\ is identical to that used
in the fits to real particle cross sections\cite{cudell,dino2}
(where $\eps _1 = 0.10$, and $\eps _2 = -0.32$ are found - the latter
for $f/A_2$ exchange)
and the value of $R$ found in the UA8 fits ($4.0 \pm 0.6$) 
is similar to the values
found in the fits to real particle cross sections, we take as a working
assumption that \sigpomtot\ is like a real particle cross section
(as is done in predicting hard diffractive cross sections\cite{is}).
Thus, $\epsilon _1$ and $\epsilon _2$ are fixed at the above values.

In Eq.~\ref{eq:tripleP2}, $|F_1(t)|^2$ is the standard 
Donnachie--Landshoff\cite{dl_elastic} 
form--factor\footnote{$F_1(t)={{4 m_p ^2 - 2.8t}
\over{4 m_p ^2 - t}}\, \x \, {1\over{(1-t/0.71)^2}}$} 
which is multiplied by a possible correction at high--$|t|$, $e^{bt}$.
Thus, the product, $|F_1(t)|^2 e^{bt}$, 
carries the \T --dependence of
$G_{{\cal P}{\cal P}{\cal P}}(t)$ and $G_{{\cal P}{\cal P}{\cal R}}(t)$
in the Mueller-Regge expansion and is assumed to be the same in both.
Physically, this means that the \pom\ has the same flux factor in
the proton, irrespective of whether the proton--\pom\ interaction proceeds
via \pom --exchange or \reg --exchange.
The products, $K \sigma_0 = G_{{\cal P}{\cal P}{\cal P}}(0)$,
and $K \sigma_0 R = G_{{\cal P}{\cal P}{\cal R}}(0)$.

The \pom\ trajectory, $\alpha (t)$ in \flux , 
was assumed to have the usual linear 
form with a quadratic term added to allow for a flattening of the
trajectory at high--$|t|$, as required by the data:
\begin{equation}
\alpha (t) \, \, = \, \, 1 + \epsilon + 0.25 t + \alpha '' t^2
\label{eq:alpha}
\end{equation}
$\epsilon$ was fixed at 0.10 in the fits.
Although we show in the next section that the effective intercept decreases 
with $s$, the only low--$|t|$ data used in the UA8 fits
was that from the lower ISR energies, where 0.10 is a good approximation.
We return to this point in Sec.~\ref{sect:traj}.

To avoid difficulties with differing experimental resolutions in
the combined ISR--UA8 data sample, simultaneous fits of Eq.~\ref{eq:tripleP2} 
were first made to data in the range $0.03 < \xi < 0.04$ and 
$|t| < 2.25$~GeV$^2$, assuming zero non--\pom --exchange background; 
then fits were made to the entire region,
$0.03 < \xi < 0.10$, including a background term of the form 
$A e ^{ct} \xi ^1$. 
All fits gave self--consistent results.
The fit values\cite{ua8dif,es1} 
of the four free parameters in Eq.~\ref{eq:tripleP2} were:
\begin{tabbing}
\hspace{5cm}\=$K\sigma _0$ \hspace{3mm}\= =\hspace{4mm}\=$0.72 \pm 0.10$ 
\hspace{5mm}\=mb GeV$^{-2}$\\
           \>$\alpha ''$   \> =\>$0.079 \pm 0.012$ \>GeV$^{-4}$\\
           \>$b$	   \> =\>$1.08 \pm 0.20$   \>GeV$^{-2}$\\
           \>$R$           \> =\>$4.0 \pm 0.6$    
\end{tabbing}
The fitted \pom\ trajectory, Eq.~\ref{eq:alpha} with $\alpha '' = 0.08$,
is shown as the shaded band in Fig.~\ref{fig:alpha}. The band edges 
correspond to $\pm 1\sigma$ error limits on $\alpha ''$.

Independent confirmation of the $\alpha (t)$ values at high--$|t|$
seen in Fig.~\ref{fig:alpha}
was obtained by fitting (resolution--smeared) Eq.~\ref{eq:tripleP2}
to the $\xi$--dependence of the UA8 data at fixed--\T\ in the {\it different}
$\xi$--region, $\xi < 0.03$, where non--\pom --exchange background could be 
ignored.
Although, in Eq.~\ref{eq:tripleP2}, the dominant 
$\xi$--dependence is in \flux\ and has the form $\xi^{1-2\alpha (t)}$, 
there are the additional (weaker) $(s')^{\epsilon} \sim \xi ^{\epsilon}$ 
dependences in the ${\cal P}{\cal P}{\cal P}$ and ${\cal P}{\cal P}{\cal R}$
terms of \sigpomtot , both of which must be included in the fit.
Because the (${\cal P}{\cal P}{\cal R}$)
term is more sharply peaked at small values of $\xi$ than is the
${\cal P}{\cal P}{\cal P}$ term, leaving it out of the 
fit\footnote{Note that this would mean assuming $R=0$, which is in blatant
disagreement with the value, $R = 4.0$, quoted above.}
causes a systematic upward shift in the resultant $\alpha(t)$. 

The solid points in Fig.~\ref{fig:alpha} show the fit values\cite{ua8dif} 
of $\alpha (t)$ at four \T -values, when both
${\cal P}{\cal P}{\cal P}$ and ${\cal P}{\cal P}{\cal R}$
terms in Eq.~\ref{eq:tripleP2} are used in the fit 
(with $R = 4.0$).
The solid points and the band in the figure are in good agreement.
The two different, but self--consistent, fits to the data
in the high--$|t|$ region give confidence in the value of the overall 
normalization constant, $K \sigma _0$, and in the \T --dependence,
$|F_1 (t)|^2 e^{bt}$. 

Table~\ref{tab:fits} summarizes the two types of fits performed by
the UA8 collaboration\cite{ua8dif} in determining $\alpha (t)$ at high \T ,
and shows which data sets were used in each. In Sect.~\ref{sect:traj},
a third independent type of 
fit is described which also yields essentially the same
results for $\alpha (t)$ at high--$|t|$ at both ISR and SPS-Collider. 

\section{An {\boldmath $s$}-dependent effective intercept}
\label{sect:sdep}
\indent

As explained above, a \sigdiftot\ prediction 
depends sensitively on the value of \eps\ used in the Flux Factor.
For each of the ISR, SPS and Tevatron$^6$ points 
in Fig.~\ref{fig:sigtot}, we have therefore found the
value of the effective \eps\ which yields the measured \sigdiftot .
Eq.~\ref{eq:tripleP2} is integrated over $\xi < 0.05$
and all \T , with the following assumptions:
\begin{itemize}
\item We assume that screening only effects the Flux Factor, \flux , and
therefore only allow the \eps\ which appears therein to change.
As stated above, our working assumption is that the proton--\pom\ total cross 
section, 
$\sigpomtot (s') = \sigma_0 [(s')^{0.10} \, + \, R \, (s')^{-0.32}]$
with $R = 4.0$, is like a real--particle total cross section and hence has 
fixed parametrization.
\item $K\sigma _0$, $R$, $b$ and $\alpha ''$ are fixed at the UA8 fit
values given in Sect.~\ref{sect:fits}. 
Although using these fixed values
is not self-consistent with allowing \eps\ to vary, 
any corrections are of order $\sim 10 \%$ and do not obscure the
essential results.
\end{itemize}

\begin{figure}
\begin{center}
\mbox{\epsfig{file=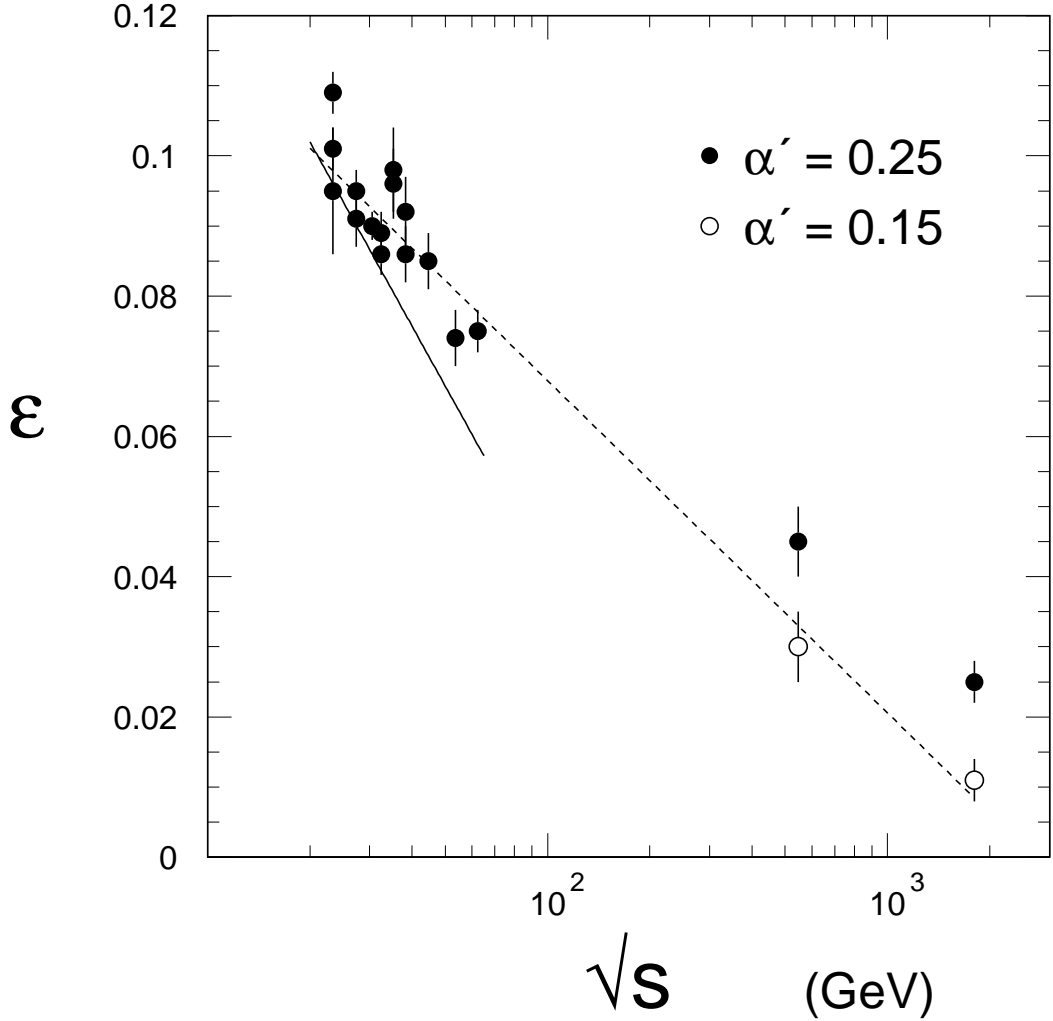,width=16cm}}
\end{center}
\caption[]{
The solid and open points are calculated effective \eps\ values 
vs. $s$, which are required to make the predicted \sigdiftot\ values 
in Fig.~\protect\ref{fig:sigtot} (solid curve)
agree with the various measured values.
To emphasize the trend, the dashed line is a fit to the solid points at 
the ISR and the open points, UA4 at the SPS-Collider and CDF at the Tevatron.
As discussed in the text, the more reliable 
solid line is the result of fitting to the ISR $d\sigma_{sd}/dt$ 
points in Fig.~\ref{fig:dsdtisr},
with \alfp\ and $\alpha ''$ also allowed to vary with $s$.
}
\label{fig:epsvss}
\end{figure}

Fig.~\ref{fig:epsvss} shows the resulting \eps\ values vs. $\sqrt{s}$; 
their errors only reflect the measurement errors in the \sigdiftot\
points.
Starting with $\epsilon = 0.10$ at the lowest end of the ISR region, the 
points display a pronounced downward trend with $s$,
reaching $\sim 0.03$ at the SPS--Collider and $\sim 0.01$ at the Tevatron.
From the fits in Sect.~\ref{sect:traj}, \alfp\ also decreases with decreasing
\eps\ (this is not surprising since the trajectory has to match up
with its $s$--independent part at higher \T ).
Since \alfp\ = 0.15 is preferred at SPS and Tevatron energies,
the dashed line in Fig.~\ref{fig:epsvss} 
is a fit to the solid points (\alfp\ = 0.25)
at the ISR and the open points at SPS-Tevatron (\alfp\ = 0.15). 
The solid line shows \eps\ vs. $s$ which results from the fits to the ISR 
data in Sect.~\ref{sect:traj}. The difference between solid points and
the solid line is that \alfp\ is not fixed at the arbitrary value of 
0.25~GeV$^{-2}$ in the latter.

\subsection{Predictions for the Tevatron}
\label{sect:cdf}
\indent

\begin{figure}
\begin{center}
\mbox{\epsfig{file=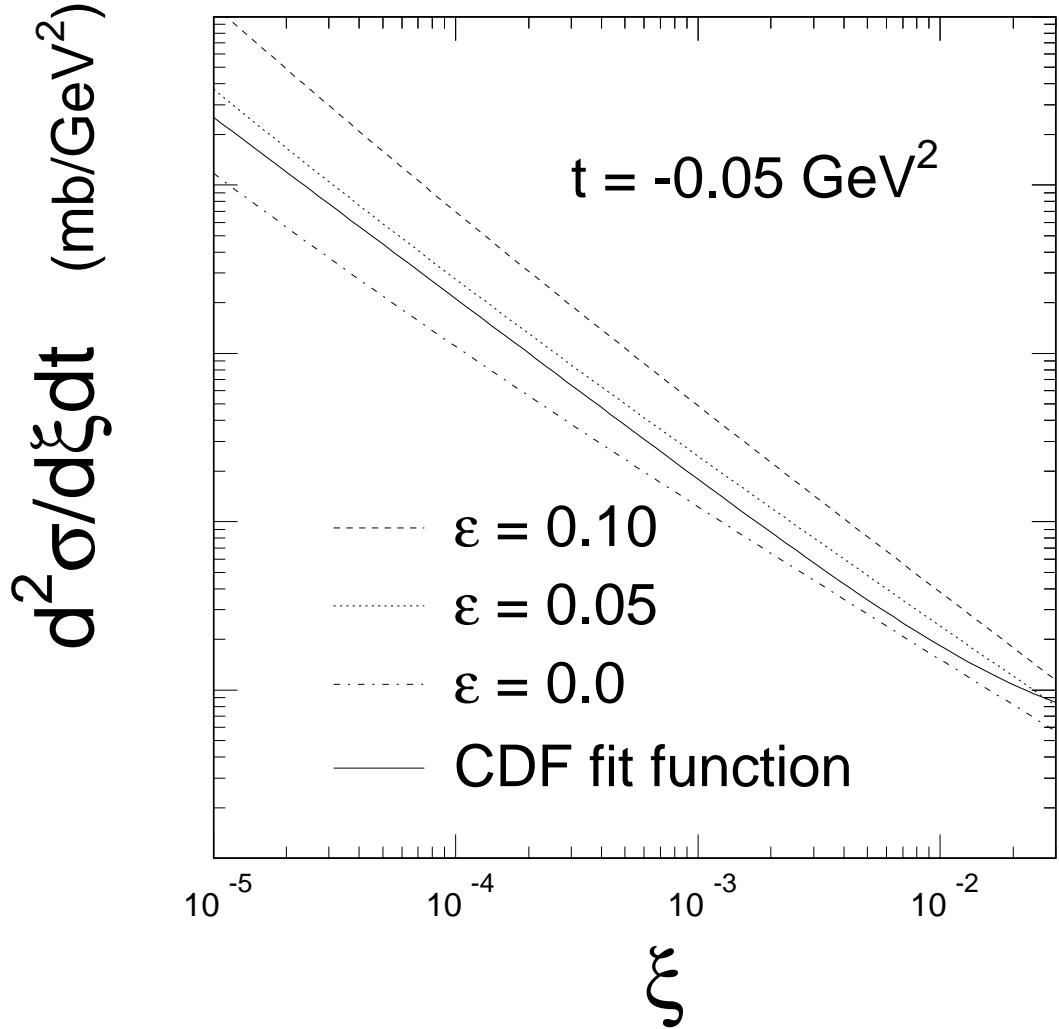,width=16cm}}
\end{center}
\caption[]{
The solid line 
is the CDF parametrization of their single diffractive differential
cross section\protect\cite{cdf} (single hemisphere)
vs. $\xi$ at $|t| = 0.05$~GeV$^2$ and \rs\ = 1800~GeV;
there is a $\pm 17\%$ quoted normalization uncertainty in \sigdiftot . 
As discussed in the text, the dashed, dotted and dash--dotted curves
show the predicted differential cross section vs. $\xi$
at three values of effective \eps\ in \flux , calculated using
Eq.~\protect\ref{eq:tripleP2}, the UA8 fit parameters and \alf\ = 0.15.
}
\label{fig:cdf}
\end{figure}

Fig.~\ref{fig:cdf} shows Eq.~\ref{eq:tripleP2} ploted vs. $\xi$ at 
\rs\ = 1800~GeV and small momentum transfer, $|t| = 0.05$~GeV$^2$, for 
several different values\footnote{$\alpha ' = 0.15$ is used for these 
curves; however, the results are almost identical for $\alpha ' = 0.25$.}
of \eps . 
The figure also shows the results of the CDF 
experiment\cite{cdf}, which they give in the form of a 
function which, after convoluting with their experimental resolution
and geometric acceptance, 
was fit to their observed differential cross 
section\footnote{For reasons explained in Sects. 5.2 and 5.3 
of Ref.~\cite{ua8dif},
we add the CDF ``signal'' and ``background'' as a good representation of their 
corrected differential cross section at small $\xi$
(after unfolding experimental resolution).}.
The CDF function (solid curve) in Fig.~\ref{fig:cdf} is seen to agree in both 
absolute magnitude and shape to a prediction with $\epsilon \sim 0.03 \pm 0.02$.
Similar results are found for the CDF function at \rs\ = 546 GeV (not shown
here). 

We have integrated these same CDF functions for $\xi < 0.05$, to obtain
$d\sigma_{sd}/dt$ at both 546 and 1800 GeV. 
At $|t| = 0.05$ Gev$^2$, both CDF values
agree with UA4\cite{ua4dif1,ua4dif2}. 
However, the CDF $|t|$-slope at 546 GeV is much steeper ($b =7.7$ GeV$^{-2}$)
than are all other ISR, SPS and Tevatron (1800 GeV) measurements. 
Since its magnitude is sufficient to account for the difference between UA4 and
CDF \sigdiftot\ at 546 GeV, we have therefore ignored that CDF point
in preparing Fig.~\ref{fig:epsvss}. 

\section{The {\boldmath \pom } trajectory}
\label{sect:traj}
\indent

From the results in Sect.~\ref{sect:sdep} at low--$|t|$, we have an 
$s$-dependent effective \pom\ intercept 
which reflects multi--\pom --exchange effects.
However, as already mentioned, at high--$|t|$ ($>1$~GeV$^2$)
the trajectory\cite{ua8dif} shows no signs of an $s$--dependence\cite{es1},
since the triple--Regge formalism describes the data between
ISR and SPS with no apparent need of damping.

In order to parametrise the full \pom\ trajectory, we resort to a somewhat
unorthodox procedure and fit the $\xi < 0.05$ integral of
Eq.~\ref{eq:tripleP2} (with free parameters to describe the effective \pom\ 
trajectory), to the \T --dependence of the total differential cross section, 
$d\sigma_{sd} /dt$ ($\xi < 0.05$). This method relies on our knowledge 
of the remaining \T --dependence, $|F_1 (t)|^2 e ^{bt}$ in \flux , which
we believe is reliable since
the results from our fits yield a \pom\ trajectory at high--\T\ which
agrees with $\alpha$ values obtained in 
the UA8 fits\cite{ua8dif} to the shape of $d\sigma / d\xi$ vs. $\xi$ 
at fixed \T\ values.

There is only one set of $d\sigma_{sd} /dt$ data above ISR energies which
covers the complete $|t|$ range from 0--2~GeV$^2$. 
Fig.~\ref{fig:dsdtsps} shows the measurements at the SPS-Collider by the UA4 
collaboration\cite{ua4dif1,ua4dif2} (open points) and by the UA8 
collaboration\cite{ua8dif} (solid points).
The UA4 data cover most of the \T --range because they come from independent
high--$\beta$ and low--$\beta$ runs at the SPS. The UA8 data only cover
the high--$|t|$ part of the range, but they are in good agreement with
the UA4 points where they 
overlap\footnote{We ignore the small ($\sim 3 \%$)
difference in \sigdiftot\ between the two \rs\ values, 
546 and 630 GeV.}. Although the poor
$\xi$ resolution of the low--$\beta$ run precludes use of the data
for fits to the $\xi$--dependence, the $d\sigma_{sd} /dt$ distribution is
hardly influenced.

\begin{figure}
\begin{center}
\mbox{\epsfig{file=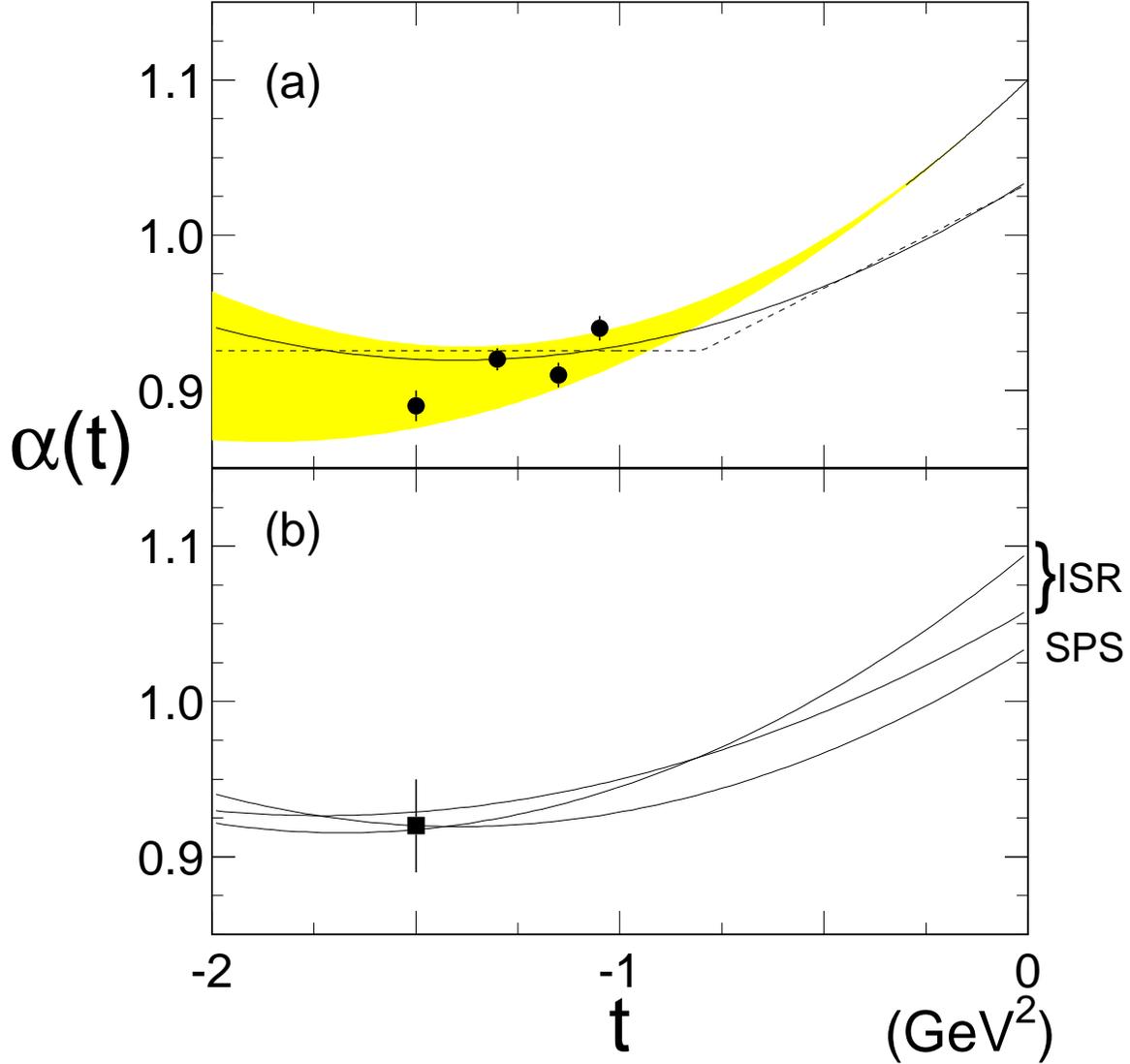,width=16.0cm}}
\end{center}
\caption[]{\pom\ Regge trajectory with $s$--dependent effective intercept;
(a) Shaded band\cite{ua8dif} is from a simultaneous fit to all 
ISR/UA8 SPS data in the range $0.03 < \xi < 0.10$;
Solid points\cite{ua8dif} are fits to the UA8 data with $\xi < 0.03$ at fixed 
\T\ (see Table~\ref{tab:fits}).
The solid and dashed curves correspond to new fits to the UA4 + UA8 SPS
$d\sigma _{sd}/dt$ with $\xi < 0.05$ as described in the text;
(b) The curve labeled ``SPS" is the same as in (a).
``ISR" curves correspond to highest and lowest ISR energies from fit to
all energies shown in Fig.~\ref{fig:dsdtisr}. The fit intercept values have
small errors (see text). The point is our best estimate of $\alpha (t)$
at $|t| = 1.5$~GeV$^2$
}
\label{fig:alpha}
\end{figure}

\begin{figure}
\begin{center}
\mbox{\epsfig{file=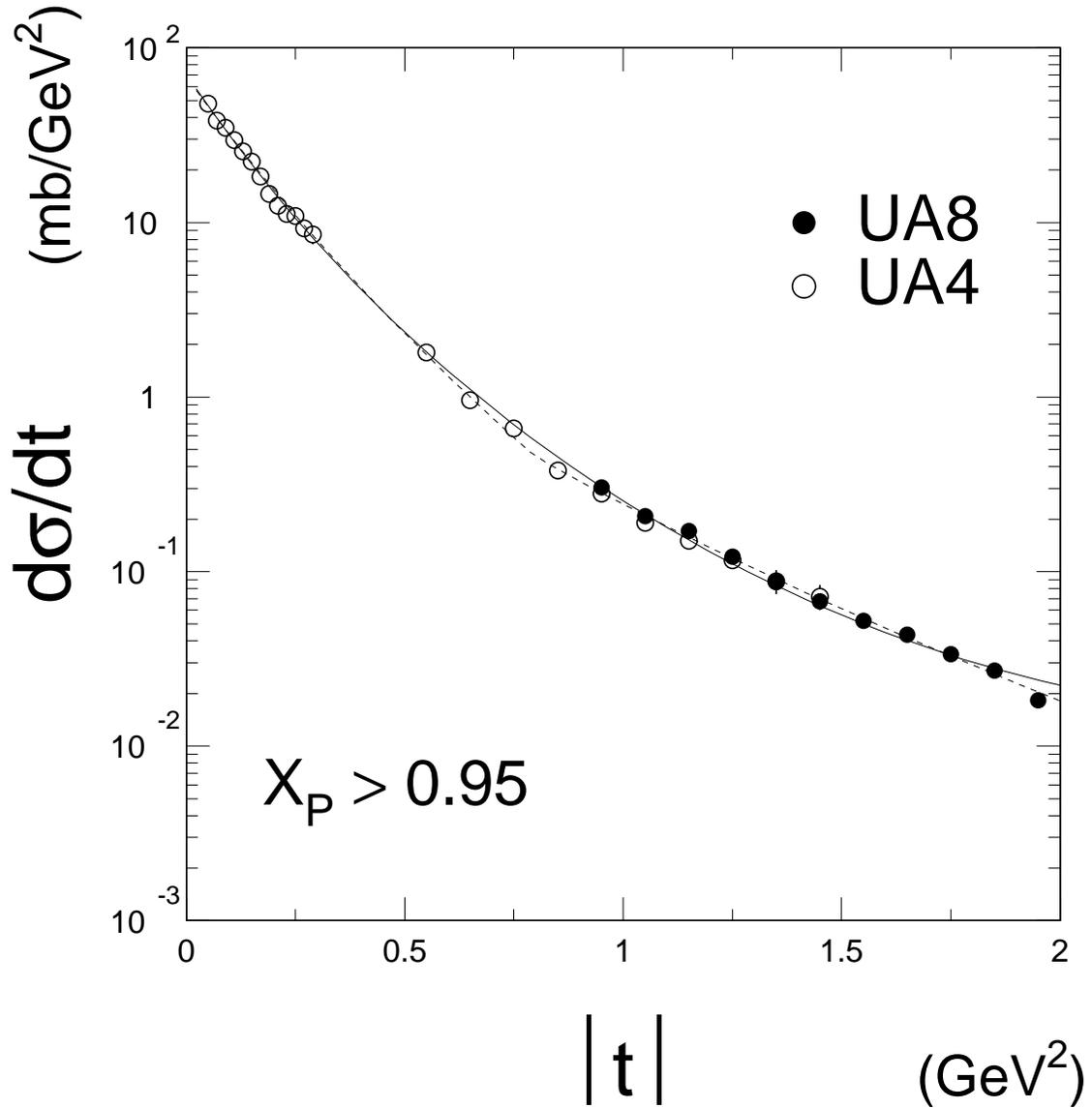,width=16cm}}
\end{center}
\caption[]{
Inclusive differential cross sections for protons (and antiprotons)
in React.~\protect\ref{eq:dif}
for $\xp > 0.95$, from experiments UA4\protect\cite{ua4dif1,ua4dif2} with 
\rs\ = 546 GeV and UA8\protect\cite{ua8dif} with \rs\ = 630 GeV).
The integral is \sigdiftot\ = $9.4 \pm 0.7$ mb.
The solid and dashed curves are fits ($\chi^2$/DF values are 4.2 and 2.0, 
respectively), corresponding to the solid and dashed
\pom\ trajectories shown in Fig.~\ref{fig:alpha}(a) and described in the text.
}
\label{fig:dsdtsps}
\end{figure}

The solid and dashed curves in Fig.~\ref{fig:dsdtsps} are 
fits (row 3 in Table~\ref{tab:fits}) of the $\xi < 0.05$ integrated 
version\footnote{In the fits, $K\sigma_0$, $b$ and $R$ are fixed at the UA8 
values given above.} 
of Eq.~\ref{eq:tripleP2} to the $d\sigma_{sd} /dt$ data points in the figure.
They correspond to the two different, but similar, parametrizations of 
$\alpha (t)$ shown in Fig.~\ref{fig:alpha}(a) (those with lower intercepts).
One is the quadratic trajectory (solid), 
$\alpha (t) = 1 + \epsilon + \alpha ' t + \alpha '' t^2$, with three
free parameters 
($0.035 \pm 0.001$, $0.165 \pm 0.002$~GeV$^{-2}$ and 
$0.059 \pm 0.001$~GeV$^{-4}$, respectively). 
The dashed trajectory consists of two straight lines, 
and also has three free parameters, the intercept, slope 
and the $|t|$ value at which the trajectory continues horizontally
to larger--$|t|$ values 
($0.033 \pm 0.001$, $0.134 \pm 0.003$~GeV$^{-2}$ 
and $0.80 \pm 0.02$~GeV$^{2}$, respectively).  
The two fitted trajectories are nearly identical. 
Remarkably, they are seen to agree with the two previous
{\it independent} determinations\cite{ua8dif} of $\alpha (t)$ 
in the high--$|t|$ region with $\xi > 0.03$.

\begin{figure}
\begin{center}
\mbox{\epsfig{file=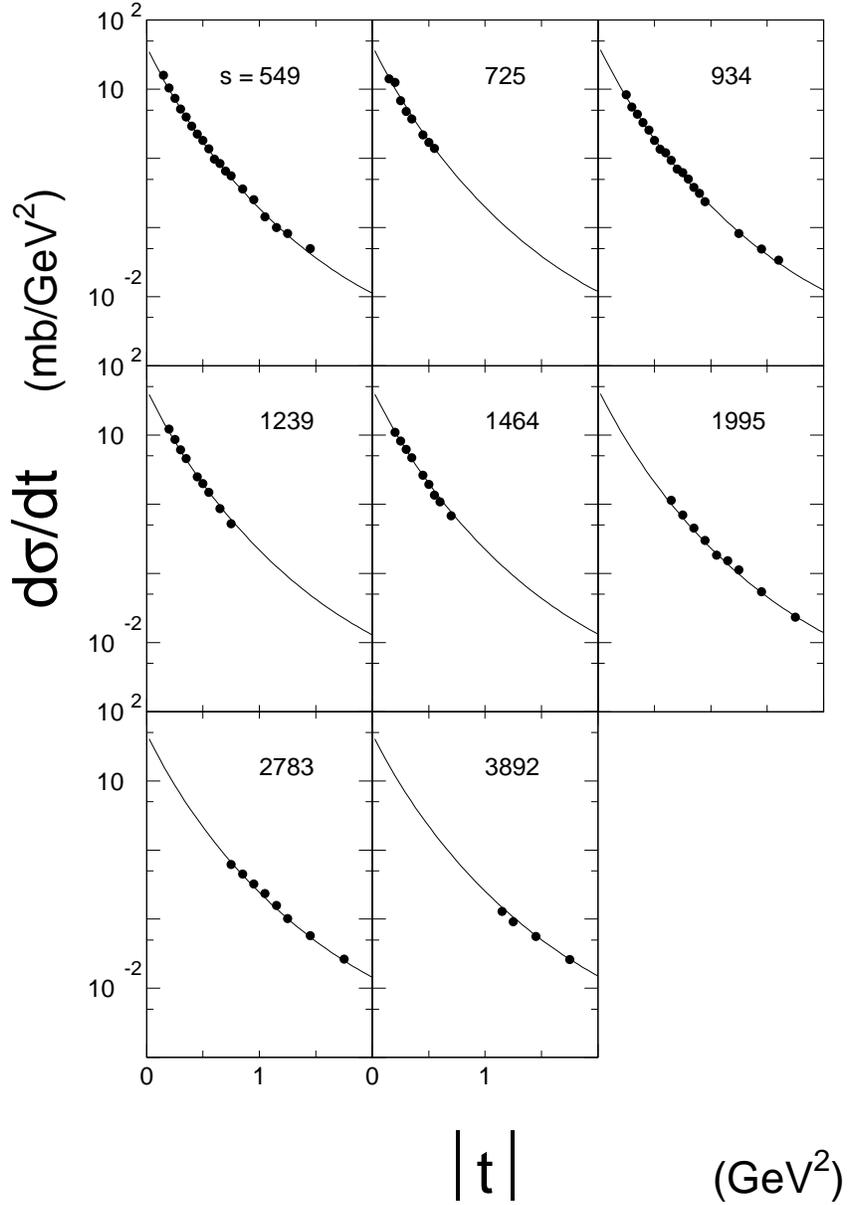,width=12cm}}
\end{center}
\caption[]{
ISR inclusive differential cross sections\cite{albrowtot,armitagetot}, 
$d\sigma_{sd}/dt$, for protons with $\xp > 0.95$ in the 
$pp$ version of React.~\protect\ref{eq:dif} (both arms) 
at the indicated $s$--values.
The solid curves result from a single 6--parameter fit of the 
$\xi < 0.05$ integrated version of Eq.~\ref{eq:tripleP2} to all data points 
shown, using $\alpha (t) = 1 + \epsilon + \alpha ' t + \alpha '' t^2$,
where each of the parameters is assumed to have an $s$--dependence of the 
type, $\epsilon (s) = \epsilon (549) + A \cdot log (s/549)$.
$\chi^2$/DF = 1.4 for the combined fit.
}
\label{fig:dsdtisr}
\end{figure}

Encouraged by this result, we turn to the corresponding ISR 
measurements\cite{albrowtot,armitagetot} of $d\sigma_{sd} /dt$ 
(see Fig.~\ref{fig:dsdtisr}).
The solid curves in Fig.~\ref{fig:dsdtisr} are the result of a single 
6--parameter fit ($\chi^2$/DF = 1.4) of the 
$\xi < 0.05$ integrated version of Eq.~\ref{eq:tripleP2} to all data points 
shown, using $\alpha (t) = 1 + \epsilon + \alpha ' t + \alpha '' t^2$.
The 6 parameters arise from assuming that $\epsilon$, $\alpha '$ and 
$\alpha ''$ each has an $s$--dependence of the type,
$\epsilon (s) = \epsilon (549) + A \cdot log (s/549)$. 
At $s = 549$~Gev$^2$, the fit parameters, $\epsilon$, $\alpha '$ and 
$\alpha ''$ are, respectively, $0.096 \pm 0.004$, $0.215 \pm 0.011$~GeV$^{-2}$ 
and $0.064 \pm 0.006$~GeV$^{-4}$. 
$\epsilon (549) = 0.10$ agrees perfectly with
the value obtained from fitting to the $s$--dependence of total cross 
sections\footnote{This justifies its use in the fits of Ref.~\cite{ua8dif}.}  
\cite{cudell,dino2} ,
while $\alpha ' (549) = 0.215$ is smaller than the conventional 0.25.  

The fit also yields the energy dependence
parameter, ``$A$'', for each of $\epsilon$, $\alpha '$ and $\alpha ''$:
$-0.019 \pm 0.005$, $-0.031 \pm 0.012$ and $-0.010 \pm 0.006$, respectively.
This allows us to plot the fitted $\epsilon$ vs $s$ in Fig.~\ref{fig:epsvss} 
(solid line) over the ISR energy range. 
The curve is more reliable than the points obtained in Sect.~\ref{sect:sdep} 
from \sigdiftot\ values, because $\alpha '$ is not fixed at the arbitrary 
value, 0.25.
The areas under the fitted curves in Figs.~\ref{fig:dsdtsps} and
\ref{fig:dsdtisr} are in good agreement with the published \sigdiftot\ values.

The effective trajectories corresponding to the fits in Fig.~\ref{fig:dsdtisr}
are plotted in Fig.~\ref{fig:alpha}(b)
at the lowest ($s = 549$~GeV$^2$) and highest ($s = 3892$~GeV$^2$) ISR 
energies. Also shown is the same SPS-Collider trajectory
as in Fig.~\ref{fig:alpha}(a). At $|t| = 1.5$~GeV$^2$, all trajectory
values agree to within about $\pm 0.01$.  
We therefore refit the ISR data of Fig.~\ref{fig:dsdtisr},
constraining all ISR trajectories to 
have the same value at $|t| = 1.5$~GeV$^2$. 
With $\chi^2$/DF = 1.4, we find
$\alpha (1.5) = 0.923 \pm 0.002$ ($\alpha (0)$ and $\alpha ' (0)$
do not change significantly). When the uncertainty in $b = 1.08 \pm 0.20$
is taken into account, the error enlarges to $\alpha (1.5) = 0.92 \pm 0.03$,
shown as the square point in Fig.~\ref{fig:alpha}(b).

\section{Discussion}
\label{sect:conclude}
\indent

Using Refs.~\cite{ua8dif,es1} and the work in the present paper,
we have seen that the triple--Regge formula 
with both ${\cal P}{\cal P}{\cal P}$ and 
${\cal P}{\cal P}{\cal R}$ terms describes all available inclusive
single--diffractive data from ISR to Tevatron, 
provided that the effective \pom\ Regge trajectory intercept, $\alpha (0)$, 
is $s$--dependent and decreases from a value, 1.10,
at low energies to a value about 1.03 at the SPS-Collider and perhaps 
smaller at the Tevatron.
The data also require\cite{ua8dif} a ``flattening" of the \pom\ 
trajectory\cite{ua8dif} at $\alpha \approx 0.92$,
for momentum transfer, $|t| > 1$~GeV$^2$.
Together, these two characteristics specify a new effective \pom\ trajectory 
in inelastic diffraction, which is 
in disagreement with the ``traditional'' soft \pom\ trajectory
obtained from fits to the energy dependence of hadronic total cross sections.
An $s$-dependent effective intercept which decreases with increasing energy
is expected from multi--\pom --exchange (screening/damping) calculations.

We find it remarkable that, despite the presence of multi--\pom --exchange 
contributions, Eq.~\ref{eq:factorhad} and the factorization of \pom\ emission 
and interaction seem to retain a high degree of validity.
This suggests that multi--\pom --exchange effects behave in an approximately 
factorizable way\cite{pumpkane}.

It is also remarkable that single--\pom --exchange with a fixed trajectory 
describes inelastic diffraction so well in the higher--$|t|$  domain,
when it is known that high--$|t|$ elastic scattering has multiple exchange
contributions\cite{dl_elastic} there. It will be useful to elaborate the
evidence for a dominant fixed \pom\ trajectory at high--$|t|$ in inelastic
diffraction:

(a) First, we note that elastic and inelastic diffraction are very 
different. There is no evidence in inelastic diffraction 
(see Figs.~\ref{fig:dsdtsps} and \ref{fig:dsdtisr}) for the characteristic
presence of the $s$--dependent dip (and break) seen in
$pp$ elastic scattering and very differently in \pap\ elastic scattering.
Indeed, we have a self--consistent set of single--\pom --exchange
fits to the $pp$ data
at the ISR (Fig.~\ref{fig:dsdtisr}) and to the \pap\ data at the
SPS--Collider (Fig.~\ref{fig:dsdtsps})

(b) The \xp\ (or $\xi$) distribution in React.~\ref{eq:dif} shows 
similar $\xp \sim 1$ peaking at high--$|t|$ ($> 1$~GeV$^2$) as at 
low--$|t|$ (see Ref.~\cite{ua8dif}), which is a signature for the flattening 
of the \pom\ trajectory at high--$|t|$. 
It is extremely interesting to note that the trajectory value we obtain,
$\alpha = 0.92 \pm 0.03$ at $|t|$ = 1.5 GeV$^2$, is consistent with
the \pom\ trajectory obtained at high--$|t|$ in $\rho ^0$ and $\phi$
photoproduction by the ZEUS collaboration\cite{zeusphoto} and
in $J/\Psi$ photoproduction by the H1 Collaboration\cite{h1photo}. 
This suggests a universal fixed \pom\ trajectory in the high--$|t|$ domain.

(c) The fits of Eq.~\ref{eq:tripleP2} 
with its embedded Regge factor, $\xi ^{1-2\alpha (t)}$, to the entire set of
differential cross section data and their $s$--dependences are highly
overconstrained and yield good results. Thus, additional complications
are not required by the data.
It is particularly impressive that the three different and independent
ways of determining $\alpha (t)$ (see Table~\ref{tab:fits})
for $|t| > 1$~GeV$^2$ in Fig.~\ref{fig:alpha} all give the same result. 
In one case, the fits\cite{ua8dif} are to all data points at ISR and 
SPS-Collider with $\xi > 0.03$ (including non--\pom --exchange background). 
Secondly, there are fits\cite{ua8dif} to the shapes of 
$d^2\sigma/d\xi dt$ vs. $\xi$ with 
$\xi < 0.03$ at fixed--\T . 
And, in the present paper, we fit to all available $d\sigma_{sd}/dt$ 
with  $\xi < 0.05$ over the entire range of \T\ at both ISR and 
SPS-Collider.
 
(d) An additional argument in favor of one--\pom --exchange at high--\T\ is 
that, in Ref.~\cite{pompom}, the extracted \pom --\pom\ total cross section,
$\sigma^{total}_{{\rm \cal P}{\rm \cal P}}$, agrees with factorization
expectations above the few--GeV mass region.

(e) Finally, the UA8 results on hard scattering\cite{ua8} at high--$|t|$ yield 
essentially the same picture of the \pom 's partonic structure as do the 
low--$|t|$ experiments at HERA and Tevatron. 

On another point, 
we understand that, naively, screening effects are expected to increase
with $|t|$. This appears to contradict our observations.
Indeed, the flattening of the \pom\ trajectory at high--$|t|$,
as well as the apparent absence of damping there\cite{es1}
(damping effects are seen to ``fade away" as $|t|$ increases from 
0.5 to 1.0 GeV$^2$) 
suggests that the trajectory is entering the perturbative domain.
For example, this change in dynamics at high--$|t|$ 
from the simple eikonal approximation could arise from the dominance
of ``small--size configurations" in the recoil nucleon\cite{fs}.
It is, in any case, 
an intriguing situation which should be given further attention.

One hopes that future calculations of multi--\pom --exchange effects will 
account for the effective intercept and slope at \T\ = 0 
which we have presented, 
as well as preserve the high degree of factorization exhibited by the data.
An additional factor which should be taken into account in
multi--\pom --exchange calculations
can be inferred from a recent result of the UA8 Collaboration on the 
analysis\cite{pompom} of double--\pom --exchange data, where both final state
observed $p$ and \ap\ are in the momentum transfer range, 
$|t| > 1.0$~GeV$^2$. 
The extracted \pom --\pom\ total cross section agrees with factorization 
expectations in the invariant mass range, $9 < \sqrt{s'} < 25$ GeV. 
However, at smaller masses, there is a pronounced enhancement of the 
\pom --\pom\ cross section, peaking in the few--GeV mass region, with about a 
factor of ten larger cross section than expected from factorization. 
Although, with a mass resolution about $\sigma = 2$~GeV, it is impossible to 
observe 
structure in the \pom --\pom\ spectrum, this result implies that there 
is at least a strong interaction in the low--mass \pom --\pom\ system, 
which can have a significant, and perhaps simplifying, 
impact on the nature of multiple \pom\ exchange. 

\section*{Acknowledgements}
\indent

We have benefited greatly from discussions with 
Alexei Kaidalov, Uri Maor and Mark Strikman on issues of damping and
screening. 
Helpful discussions with John Dainton are also appreciated. 
We also wish to thank the CERN laboratory, where much of this work
was done, for their long hospitality.


\begin{table}
\centering
\begin{tabular}{||c|c|c||c|c||c|c||}                         
\hline
Results &Type of &Cross  &\multicolumn{2}{|c||}{SPS} 
                                          &\multicolumn{2}{|c||}{ISR}\\
in Fig.~\protect\ref{fig:alpha} &fit to data &section &low--$|t|$ &high--$|t|$  
&low--$|t|$ &high--$|t|$ \\
        &       &used ? &&                          &&\\
\hline
 &     &      &         &     &      &     \\
Points & Fits to $d\sigma/d\xi$ &NO&&$\surd$&&\\
       & at fixed-\T\ ($\xi < 0.03$)               &&&&&\\
 &     &      &         &     &      &     \\
Shaded   &Fits to $d^2 \sigma_{sd} /d \xi dt$ &YES& &$\surd$ &$\surd$ &$\surd$\\
Band     &($0.03 < \xi < 0.10$)  &&&&&\\
 &     &      &         &     &      &     \\
Curves &Fit to $d\sigma_{sd}/dt$ ($\xi < 0.05$) &YES &$\surd$ &$\surd$ &&\\
 &     &      &         &     &      &     \\
Curves &Fit to $d\sigma_{sd}/dt$ ($\xi < 0.05$) &YES & & & $\surd$ &$\surd$\\
&&&&&&\\
\hline
\end{tabular}
\caption[]{
The three types of fits which yield self-consistent \pom\ Regge trajectories
in the $|t| > 1$~GeV$^2$ domain, shown in Fig.~\ref{fig:alpha}.
The fits\cite{ua8dif} labeled ``shaded band" include 
non--\pom --exchange background.
}
\label{tab:fits}
\end{table}

\clearpage
 
\clearpage

\clearpage

\clearpage
 
\clearpage
 
\clearpage
 
\clearpage
 
\end{document}